\documentclass[12pt]{iopart}

\usepackage{amssymb}
\usepackage{graphicx}

\usepackage[ps2pdf,colorlinks=true, pagebackref=false, bookmarks=true,bookmarksopen=true,bookmarksnumbered=true]{hyperref}

\newcommand{\Cu}{\ensuremath{\mathrm{Cu}}}
\newcommand{\Ni}{\ensuremath{\mathrm{Ni}}}
\newcommand{\Nb}{\ensuremath{\mathrm{Nb}}}
\newcommand{\Al}{\ensuremath{\mathrm{Al}}}
\newcommand{\Si}{\ensuremath{\mathrm{Si}}}
\renewcommand{\O}{\ensuremath{\mathrm{O}}}
\newcommand{\F}{\ensuremath{\mathrm{F}}}
\renewcommand{\S}{\ensuremath{\mathrm{S}}}
\newcommand{\Ar}{\ensuremath{\mathrm{Ar}}}

\usepackage{xspace}
\newcommand{\units}[1]{\ensuremath{\,\mathrm{#1}}\xspace}

\newcommand{\Section}[1]{}
\newcommand{\Subsection}[1]{}
\newcommand{\W}[1]{\textcolor{black}{#1}}

\begin{document}

\title{%
  Critical current diffraction pattern of SIFS Josephson junctions with step-like F-layer
}

\author{M. Weides$^{1,2}$, U. Peralagu$^{1,3}$, H. Kohlstedt$^{1,4}$, J. Pfeiffer$^5$, M. Kemmler$^5$, C. G\"urlich$^5$, E. Goldobin$^5$, D. Koelle$^5$, R. Kleiner$^5$}
\ead{weides@physics.ucsb.edu}

\address{$^1$%
  Institute of Solid State Research, Research Centre, J\"ulich, 52425 J\"ulich, Germany
}
\address{$^2$%
Present address: Department of Physics, University of California,
  Santa Barbara, CA 93106, USA}
\address{$^3$%
Present address:
Department of Electronics and Electrical Engineering,
  University of Glasgow, Glasgow G12 8LT, UK}

\address{$^4$%
Present address:
 Nanoelektronik, Christian-Albrechts-Universit\"at zu Kiel, Kaiserstra{\ss}e 2, 24143 Kiel, Germany}

\address{$^5$%
  Physikalisches Institut-Experimentalphysik II and Center for Collective Quantum Phenomena,   Universit\"at T\"ubingen,
  Auf der Morgenstelle 14,
  72076 T\"ubingen, Germany
}

\date{\today}

\begin{abstract}
  We present the latest generation of superconductor-insulator-ferromagnet-superconductor Josephson tunnel junctions with a step-like thickness of the ferromagnetic (F) layer. The F-layer thicknesses $d_1$ and $d_2$ in both halves were varied to obtain different combinations of positive and negative critical current densities $j_{c,1}$ and $j_{c,2}$. The measured dependences of the critical current on applied magnetic field can be well described by a model which takes into account different critical current densities (obtained from reference junctions) and different net magnetization of the multidomain ferromagnetic layer in both halves.
\end{abstract}


%\keywords{%
%  Josephson junctions, $\pi$-junction, Superconductor ferromagnet superconductor junctions
%}%Use showkeys class option if keyword display desired

\maketitle

\Section{Introduction}

%what are 0 and pi

Superconducting spintronic elements, made up by superconducting (S) and ferromagnetic (F) layers, may improve future classical and quantum computing devices \cite{KatoPRB07}.
The underlying physics is based on the difference in spin-order (antiparallel alignment in S and parallel in F): the Cooper pair singlet injected into a ferromagnet gains a finite center of mass momentum, which, in turn, leads to an oscillating phase of the superconducting order-parameter \cite{buzdin05RMP}. In the ground state, the conventional, e.g. SIS (I: tunnel barrier), Josephson junctions (JJs) have the same sign of the order parameter in both superconducting electrodes, whereas in SFS or SIFS JJs the order parameters may have opposite signs \cite{VasenkoPRB,OboznovRyazanov06IcdF,WeidesHighQualityJJ,Bannykh08,SprungmannCoFe09}, i.e., one deals with a ``$\pi$ JJ'', as the Josephson phase is $\phi=\pi$ in the ground state due to a negative critical current $I_c < 0$. The conventional JJs with $I_c > 0$ are called ``0 JJ'' in this context, as they have $\phi=0$ in the ground state. Generally speaking $I_c$ is an oscillating and decaying function of the F-layer thickness $d_F$, see Fig.~\ref{Fig:SketchopiJJ}(a). By using an F-layer with a step-like thickness, $d_1$ in one half and $d_2$ in the other half, as shown in Fig.~\ref{Fig:SketchopiJJ}(b), one may fabricate a JJ with different $I_{c,1}=I_c(d_1)$ in one part of the JJ and $I_{c,2}=I_c(d_2)$ in the other part. If $I_{c,1}$ and $I_{c,2}$ have different signs, one obtains a $0$--$\pi$ JJ \cite{WeidesFractVortex}, in which 0 and $\pi$ ground states compete with each other. At certain conditions \cite{Bulaevskii:0-pi-LJJ} the ground state of the JJ is double degenerate, corresponding to a vortex of supercurrent circulating either clock- or counterclockwise, and thereby creating a spontaneous magnetic flux \cite{Goldobin02SF} $|\Phi|\lesssim\Phi_0/2$, where $\Phi_0$ is a magnetic flux quantum.

%old gen of 0-pi
Various $0$-$0$ ($I_{c,1},I_{c,2}>0$), $\pi$-$\pi$ ($I_{c,1},I_{c,2}<0$) and $0$--$\pi$ ($I_{c,1}>0,I_{c,2}<0$) SIFS JJs based on diluted $\Ni\Cu$ alloy were already fabricated and studied by us \cite{WeidesFractVortex,WeidesSteppedJJ,WeidesIEEE,KemmlerPRB10}. In comparison with other types of 0-$\pi$ JJs, such as SFS or d-wave/s-wave \cite{Smilde:ZigzagPRL,Ariando:Zigzag:NCCO}, the intrinsic capacitance of SIFS JJs makes these JJs underdamped, so that one can study the dynamics, make spectroscopy \cite{PfeifferPRB08} or use them in macroscopic quantum circuits \cite{KatoPRB07}.

In this paper we present a new generation of SIFS ($\Nb|\Al\O_x|\Ni\Cu|\Nb$)  JJs with a step in F-layer thickness. In comparison with the previous process the new one provides more superior JJs.
First, additional $\Al$ interlayers within the  bottom $\Nb$ electrode decrease the SI interface roughness. Second, computer control of the F-layer step etch enhances reproducibility. Third, a lower F-layer gradient gives a better sample yield \cite{F-layer}.\\
\W{The multilayer is deposited by DC magnetron sputtering on thermally oxidized 4-inch $\Si$ substrates. The $160\:\rm{nm}$ thick Nb bottom electrode, made up by four $40\:\rm{nm}$ Nb layers, each separated by $2.4\:\rm{nm}$ $\Al$ layers to reduce roughness, was covered by a $5\:\rm{nm}$ thick Al layer and thermally oxidized for $30\:\rm{min}$ at $10^{-2}\:\rm{mbar}$ residual oxygen pressure and room temperature. To obtain many structures with different
F-layer thicknesses in one fabrication run, we deposit a wedge-shaped F-layer (i.e. $\Ni\Cu$) alloy in order to minimize inevitable run-to-run variations. The F-layer gradient was reduced by a factor of $\sim 2$ compared to the previous process \cite{WeidesSteppedJJ}. The multilayer stack was covered with a $40\:\rm{nm}$ $\Nb$ cap layer.\\
The stepped junctions were patterned using a four-level optical photolithographic mask procedure including $\S\F_6$ reactive etching and $\Ar$ ion-beam milling. The junctions were partly protected with photoresist to define the step location in the F-layer, followed by i) selective $\Nb$ removal via $\S\F_6$ reactive etching, ii) $\Ar$ ion-etching of the $\Ni\Cu$ and iii) in-situ deposition of $\Nb$. The insulating layer between top and bottom electrode is self-aligned by ion-beam etching down to the $\Al\O_x$ tunnel barrier and anodic oxidation of the bottom $\Nb$ electrode. Finally the top wiring is deposited.\\
Various junctions were placed on the wafer within a narrow row perpendicular to the gradient in the F-layer thickness and were replicated along this gradient. One row contained a triplet of junctions including:
\begin{itemize}
\item reference JJ with etched, uniform F-layer of thickness $d_1$
\item reference $\pi$ JJ with as deposited, uniform F-layer of thickness $d_2$
\item stepped $0\mbox{-}\pi$ JJ with step $\Delta d_F$ in the F-layer thickness from $d_1$ to $d_2$
\end{itemize}}

Our data are compared with a simple model, which takes into account different $I_{c,1} \neq I_{c,2}$ and different net magnetizations $M_1 \neq M_2$ in each half. Both the multidomain structure of the F-layer, yielding a stochastically distributed local magnetization, and the difference in magnetic thickness in both halves result in $M_1 \neq M_2$.

\Section{Model}

We consider a SIFS JJ, shown in Fig.~\ref{Fig:SketchopiJJ}(b) with $I_{c,1}\neq I_{c,2}$. The fluxes $\Phi_{M,1} \neq \Phi_{M,2}$, created by in-plane F-layer net magnetizations $M_1$ and $M_2$, are added to the flux $\Phi$ provided by an external uniform magnetic field $H$. \W{Note that $M_1$ and $M_2$  can be independently aligned either parallel or antiparallel to the applied field. Thus, their sign can be either positive (parallel alignment) or negative (antiparallel alignment).} The restriction to 1D magnetization is experimentally justified for weak magnets, the more general case of 2D in-plane magnetization in SIFS JJs is discussed elsewhere \cite{WeidesAnisotropySIFS}.
In the following we use an index $i=1,2$ to refer to two different parts of our JJ. The magnetic flux density is $b_{i}= \mu_0 H2\lambda_L+\Phi_{M,i}/L_{i}$ with the London penetration depth $\lambda_L$ and the length $L_i$ of each part.
\\
For a short JJ of length $L=L_1+L_2$ and width $w$, both $\lesssim 4\lambda_J$ ($\lambda_J$ is the Josephson penetration length), the local phase is $\phi_{i}(x)=\phi_0+2\pi/\Phi_0 \cdot b_{i} x $ with an arbitrary initial phase $\phi_0$ and the maximum Josephson supercurrent for each $H$ is given by
\begin{equation}
  I_c(H) =
  \max_{\phi_0} \left[ \sum\limits^{2}_{i=1} \left( I_{c,i}\int\limits_{L_i} \sin{\phi_{i}(x)}\,dx \right) \right] \frac{1}{L}
  %=\frac{I_{c,1}}{L/2}\int\limits^{0}_{-L/2} \sin{\phi_{1}(x)}\,dx
  %+\frac{I_{c,2}}{L/2}\int\limits^{L/2}_{ 0} \sin{\phi_{2}(x)}\,dx
  ,\label{Eq:Ic(H)}
\end{equation}
where $I_{c,i}=j_{c,i} w L$. Further we assume that $L_1=L_2=L/2$.%, where $I_{c,i}=j_{c,i} w L$.

The effect of the input parameters, such as $I_{c,i}$ and $\Phi_{M,i}$, on $I_c(H)$ was discussed for our first $0$--$\pi$ junction ($I_{c1}>0$, $I_{c2}<0$) in Ref.~\cite{KemmlerPRB10}. Now, using our latest generation of samples, we check the applicability of this model in the whole range of $I_{c,2}/I_{c,1}=-1\ldots+1$.

\Section{Samples}
\iffalse
\begin{figure}[tb]
  \begin{center}
    \includegraphics[width=8.6cm]{Sketch0piJJ.eps}
  \end{center}
  \caption{(Color online)
    (a) The dependence $|I_c(d_F)|$ ($(d_F)$ as deposited initially) for JJs with non-etched (stars) and
    etched (circles) F-layer. The symmetric $0$--$\pi$ junction (solid line) has $I_c(d_1)=-I_c(d_2)$.
    (b) Sketch of a step-type  $0$--$\pi$  JJ and its local phase $\phi_{1,2}(x)$. $H$ and $\Phi_{M,i}$ are orientated along the same axis.
    }
  \label{Fig:SketchopiJJ}
\end{figure}
\fi

Fig.~\ref{Fig:SketchopiJJ}(a) shows the experimentally measured $|I_c(d_F)|$ dependence for reference JJs without an F-layer step. For JJs with an as-deposited (non etched) F-layer the $I_c(d_F)$ vanishes at $d_F\approx6.8\units{nm}$ due to crossover from 0 to $\pi$ ground state, i.e. the sign change of $I_c(d_F)$. The etched away F-layer thickness $\Delta d_F\approx1.7\units{nm}$ was determined by the shift of $|I_c(d_F)|$ dependence along $d_{F}$ axis after etching.
Note that we have two reference JJs for each $d_F$: the one was etched ($d_F=d_1$) and another left as deposited ($d_F=d_2$).

\iffalse
\begin{figure}[tb]
  \begin{center}
    \includegraphics[width=8.6cm]{IcH2dSimu.eps}
  \end{center}
  \caption{(Color online)
    Surface plot of $I_c(H)$ for $I_{c,2}/I_{c,1}=-1\ldots1$ and fluxes $\Phi_{M,i}=0$ calculated by Eq. \ref{Eq:Ic(H)}. The measured $I_c(H)$ pattern (right scale, dotted line: baseline) were shifted by their $I_{c,2}/I_{c,1}$ ratio denoted in the black box (left scale).
  }
\label{Fig:IcH2dSimu}
\end{figure}
\fi

For JJ with a step in the F-layer, Eq.~\ref{Eq:Ic(H)} predicts the $I_c(H)$ patterns shown as color plot in Fig.~\ref{Fig:IcH2dSimu} for $I_{c,2}/I_{c,1}=-1\ldots1$. The full range $I_{c,2}/I_{c,1}=-\infty...\infty$ is included by an index change. The fluxes $\Phi_{M,i}$ were set to zero. Experimentally the $I_c(H)$ dependences are measured at $4.2\units{K}$ for three JJs with $I_{c,2}/I_{c,1}<0$ ( $0$--$\pi$  JJs) and for four JJs with $I_{c,2}/I_{c,1}>0$ ($0$-$0$ or $\pi$-$\pi$ JJs). They were taken on zero-field cooled samples for both field sweeping direction. The magnetic field was applied in-plane and parallel to the F-layer step, see Fig. \ref{Fig:SketchopiJJ} (b). The measured curves are shifted in Fig.~\ref{Fig:IcH2dSimu} according to their $I_{c,2}/I_{c,1}$ ratio. At zero magnetic field the $I_c(0)$ changes from a global maxima at $I_{c,2}/I_{c,1}=1$ to a global minimum at $I_{c,2}/I_{c,1}=-1$, as the integrals in Eq.~\ref{Eq:Ic(H)} gradually cancel out. For applied magnetic flux equal to the \emph{odd} multiples of $\Phi_0$ the supercurrents cancel out completely only for the JJ with uniform $I_c$ ($I_{c,2}/I_{c,1}=1$). For \emph{even} multiples of $\Phi_0$ the $I_c=0$ results from cancelation of supercurrents separately in each part of the JJ, irrespectively of the $I_{c,2}/I_{c,1}$ ratio. Thus, Fig.~\ref{Fig:IcH2dSimu} confirms the \emph{qualitative} agreement of experimental data with the theoretical model taking into account $I_{c,1}\neq I_{c,2}$.

\iffalse
\begin{figure}[tb]
  \begin{center}
    \includegraphics[width=8.6cm]{IcH}
  \end{center}
  \caption{(Color online)
    Measured $I_c(H)$ (voltage criteria of $0.5\units{\mu V}$) of $0$, $\pi$ and  $0$--$\pi$  JJs for three JJ sets.  In (a) and (b) the  $0$--$\pi$  JJs have $I_{c,1}\neq -I_{c,2}$ (dip at $H\approx 0$ is not fully developed). The asymmetry of the main maxima indicates a difference in local magnetizations $\Phi_{M,i}$. In (c) the  $0$--$\pi$  JJ is symmetric, i.e. $I_{c,1}=-I_{c,2}$. The insets depict the central dips of the asymmetric  $0$--$\pi$  JJs. Calculations (a)--(c) were done with $\Phi_{M,i}=0$  (dashed lines) and fitted $\Phi_{M,i}$ (solid lines).
  }
  \label{Fig:IcH}
\end{figure}
\fi

In Fig.~\ref{Fig:IcH} we show that a good \emph{quantitative} agreement can be obtained by also allowing for independent magnetizations in both halves, i.e., $\Phi_{M,1}\neq\Phi_{M,2}$. Although our analytical model can be applied to $0$-$0$,  $\pi$-$\pi$ or  $0$--$\pi$  JJs, we focus on the $0$--$\pi$  JJs as they are  more sensitive to parameters and more interesting for applications. Three sets of JJs have the F-layer thicknesses $d_2=7.3$, $7.68$ and $8.13\units{nm}$. Their lateral sizes ($100\times50\units{\mu m^2}$) were comparable or smaller than $\lambda_J$. The longest sample ($0$ JJ of set $d_2=7.3\units{nm}$) has $\lambda_J\approx75\units{\mu m}$. The measured $I_c(H)$ of the  $0$--$\pi$  and the 0 and $\pi$ reference JJs are shown in Fig.~\ref{Fig:IcH} together with the calculated pattern for the  $0$--$\pi$  JJs.

The magnetic field dependencies $I_{c,1}(H)$ and $I_{c,2}(H)$ of the reference junctions are nearly ideal Fraunhofer patterns with oscillation period $\mu_0\Delta H \approx 93\units{\mu T}$, see Fig.~\ref{Fig:IcH}. The small shifts along the $H$-axis are attributed to a net magnetization of the F-layer. For the  $0$--$\pi$  JJs the $I_c(H)$ oscillation period was $\approx 184\units{\mu T}$, nearly twice as large as for the reference JJs, as expected from theory.
The plateau with a weakly developed dip in Fig.~\ref{Fig:IcH}(a) and partially developed dip in Fig.~\ref{Fig:IcH}(b) at $H\approx 0$ are caused by the critical current asymmetry, i.e., $I_{c,1}\neq -I_{c,2}$. This can be seen from the calculated $I_c(H)$ pattern, too. For both samples the asymmetry of the main maxima and the shift of the $I_c(H)$ pattern along the $H$-axis indicate a difference in the net magnetizations $\Phi_{M,i}$. The best fit of the experimentally measured curves using Eq.~(\ref{Eq:Ic(H)}) with $\Phi_{M,i}$ as fitting parameters  yields the fluxes $\Phi_{M,i}<0.2\Phi_0$, i.e., dividing them by an area $6\units{nm} \times 50\units{\mu m}$ of the F-layer in one half of the junction we obtain the net magnetization $<1.4\units{mT}$, whereas for a fully polarized $\Ni\Cu$ alloys the saturation magnetization of $100\units{mT}$ was reported \cite{AhernNiCu}. The fitting parameter gives the mean magnetization in each half, indicating that the zero field cooled F-layer is in a multi-domain state. We attribute the remaining discrepancy with data to local nonuniformity of magnetization. Another feature --- bumped minima in $I_c(H)$ for $I_{c,1}\neq -I_{c,2}$ and $\Phi_{M,1}\neq \Phi_{M,2}$ \cite{KemmlerPRB10}--- appear very close to the measurement resolution limit and have not been further investigated. The $I_c(H)$ dependence of the sample with $d_2=8.13\units{nm}$ shown in Fig.~\ref{Fig:IcH}(c) is very symmetric, indicating $\Phi_{M,1} = \Phi_{M,2} = 0$. The maximum critical currents at the left and right maxima of the $I_c(H)$ pattern for the  $0$--$\pi$  JJ are $99.2$ and $98.9\units{\mu A}$, respectively. They differ by less than $1\%$, and were $\approx 0.72 I_{c,i}$, as expected from the theory \cite{Smilde:ZigzagPRL}. The central feature of $0$--$\pi$  JJs --- the dip or plateau at zero field with critical current given by $|I_c(0)/I_{c,1}| = \left(1-|I_{c,2}/I_{c,1}|\right)/2$ --- is fairly affirmed for all three $0$--$\pi$  JJs.

\Section{Summary}

In summary, we have fabricated SIFS JJs with and without a step in the F-layer thickness using state-of-the art SIFS JJ technology and produced $0$--$0$, $0$--$\pi$, $\pi$--$\pi$ JJ with different $I_c$ in the two parts. The experimentally measured $I_c(H)$ dependences can be well described by the model assuming different $j_c$ and different net magnetization in different parts of F-layer. The fits are much better than for the previous generation of samples \cite{KemmlerPRB10}. The presented results demonstrate a good understanding of the physics of such  $0$--$\pi$  JJs. \\Thus the presented SIFS technology may provide the JJs with step-wise tailored $j_c(x)$ including the $\pi$ regions with $j_c(x)<0$.

\Section{Acknowledgements}
Support via DFG (SFB/TRR-21, WE 4359/1-1) and AvH foundation (M.W.) is gratefully acknowledged.

\iffalse
\bibliographystyle{apsprl}
\bibliography{C:/Users/Martin/Juelich/eigene_Paper/BIB,this}

\begin{thebibliography}{10}
\bibitem{F-layer}
The F-layer thicknesses were determined by XRD, compared to profiler
  measurement in
  \cite{WeidesFractVortex,WeidesSteppedJJ,WeidesIEEE,KemmlerPRB10}.
\end{thebibliography}
\fi

\newpage

\newpage

\begin{figure}[tb]
  \begin{center}
    \includegraphics[width=8.6cm]{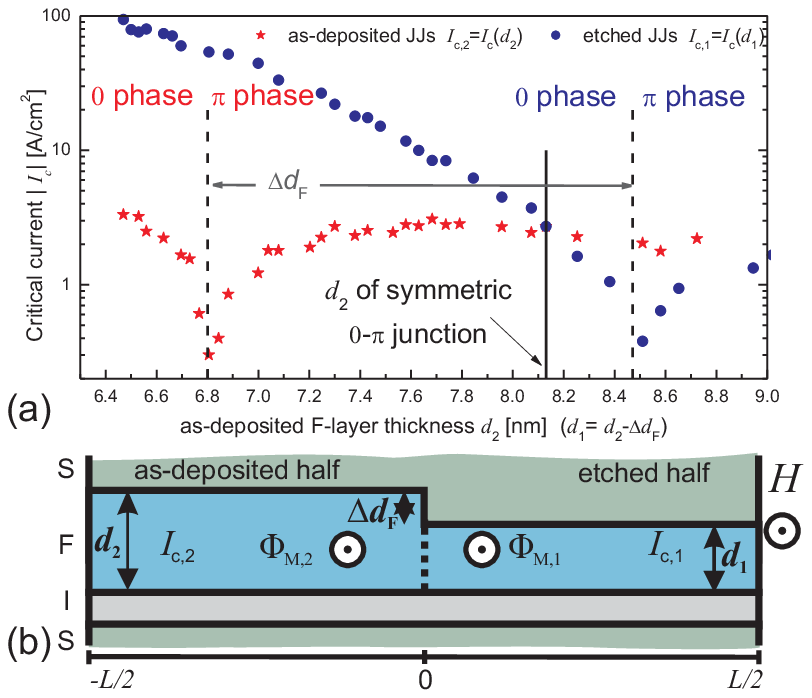}
  \end{center}
  \caption{(Color online)
    (a) The dependence $|I_c(d_F)|$ ($(d_F)$ as deposited initially) for JJs with non-etched (stars) and
    etched (circles) F-layer. The symmetric  $0$--$\pi$  junction (solid line) has $I_c(d_1)=-I_c(d_2)$.
    (b) Sketch of a step-type  $0$--$\pi$  JJ and its local phase $\phi_{1,2}(x)$. $H$ and $\Phi_{M,i}$ are orientated along the same axis.
    }
  \label{Fig:SketchopiJJ}
\end{figure}

\newpage
\begin{figure}[tb]
  \begin{center}
    \includegraphics[width=8.6cm]{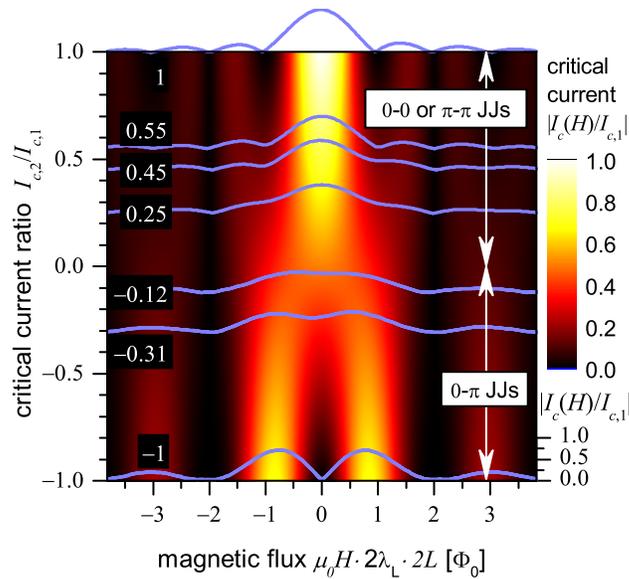}
  \end{center}
  \caption{(Color online)
    Surface plot of $I_c(H)$ for $I_{c,2}/I_{c,1}=-1\ldots1$ and fluxes $\Phi_{M,i}=0$ calculated by Eq. \ref{Eq:Ic(H)}. The measured $I_c(H)$ pattern (right scale, dotted line: baseline) were shifted by their $I_{c,2}/I_{c,1}$ ratio denoted in the black box (left scale).
  }
\label{Fig:IcH2dSimu}
\end{figure}

\newpage
\begin{figure}[tb]
  \begin{center}
    \includegraphics[width=8.6cm]{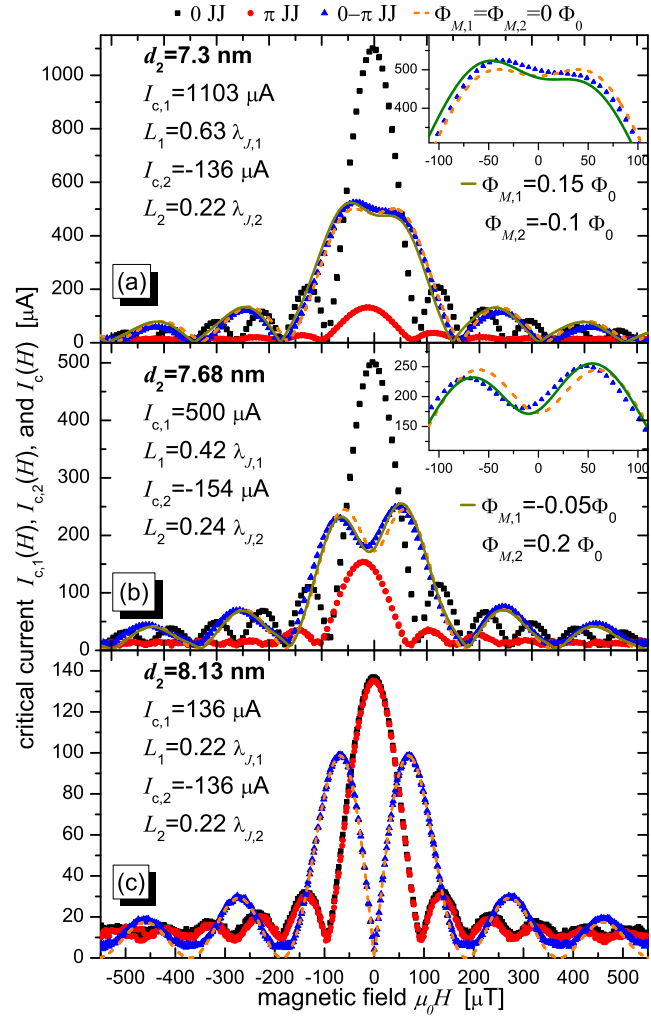}
  \end{center}
  \caption{(Color online)
    Measured $I_c(H)$ (voltage criteria of $0.5\units{\mu V}$) of $0$, $\pi$ and  $0$--$\pi$  JJs for three JJ sets.  In (a) and (b) the  $0$--$\pi$  JJs have $I_{c,1}\neq -I_{c,2}$ (dip at $H\approx 0$ is not fully developed). The asymmetry of the main maxima indicates a difference in local magnetizations $\Phi_{M,i}$. In (c) the  $0$--$\pi$  JJ is symmetric, i.e. $I_{c,1}=-I_{c,2}$. The insets depict the central dips of the asymmetric  $0$--$\pi$  JJs. Calculations (a)--(c) were done with $\Phi_{M,i}=0$  (dashed lines) and fitted $\Phi_{M,i}$ (solid lines).
  }
  \label{Fig:IcH}
\end{figure}

\end{document}